\documentclass[journal]{IEEEtran}

\usepackage{helvet}

\usepackage{amsthm}
\usepackage[cmex10]{amsmath}
\usepackage{amsfonts}
\usepackage{amssymb}
\usepackage{longtable}
\ifCLASSINFOpdf
\usepackage[pdftex]{graphicx}
\usepackage{subfigure}
\usepackage{multirow}
\else

\usepackage{graphicx}
\graphicspath{ {./figures/} }
\usepackage{subfigure}
\usepackage{amssymb}

\fi
\usepackage{epstopdf}
\usepackage{color} 

\usepackage[font=normalsize]{caption}
\usepackage[table,xcdraw]{xcolor}
\usepackage{algorithmic,algorithm}

\usepackage[numbers,sort&compress]{natbib}
\usepackage{color}

\usepackage{url}

\usepackage{mdwtab}

\renewenvironment{IEEEbiography}[1]
  {\IEEEbiographynophoto{#1}}
  {\endIEEEbiographynophoto}
  
\begin{document}	
	
	\title{Radio Radiance Field: The New Frontier of Spatial Wireless Channel Representation}

\author{Haijian Sun,~\IEEEmembership{Senior Member, IEEE,} and 
		Feng Ye,~\IEEEmembership{Senior Member,~IEEE}	%
		
		\thanks{Haijian Sun is with the School of Electrical and Computer Engineering, University of Georgia, Athens, GA, USA. E-mail: hsun@uga.edu.}
		\thanks{Feng Ye is with the Department of Electrical and Computer Engineering, University of Wisconsin-Madison, Madison, WI, USA. E-mail: feng.ye@wisc.edu.} 
        \thanks{The work of H. Sun is supported in part by the National Science Foundation under Award ECCS-2521925 and Award CNS-2236449.}
        \thanks{The work of F. Ye is supported in part by the National Science Foundation under Award ECCS-2336234.}%
}%
	
		\maketitle

\maketitle

\begin{abstract}

Massive MIMO, among other ground-breaking technologies, is being developed for the next-generation wireless systems to support requirements in terms of data rates, reliability, latency, intelligence, security and energy efficiency. 
Accurate channel estimation remains a key challenge in fully exploiting massive MIMO. While recent research has explored aspects such as near-field effects, spatial non-stationarity, and channel sparsity, many practical estimation and modeling techniques still provide limited CSI, often dominated by aggregate channel gain and delay, without full spatial characteristics.
Although wideband models and phased-array techniques can capture delay and angular information, many practical estimation methods still lack comprehensive spatial resolution, including polarization, which limits their effectiveness for advanced massive MIMO techniques.
This article introduces the concept of radio radiance field (RRF), which captures the spatial distribution and directionality of radio propagation. From RRF, a comprehensive spatial representation of the wireless channel, referred to as Spatial-CSI, can be derived.
Owing to the comprehensive geometric and radio information, RRF can be implemented directly for beamforming, delay-alignment modulation, and many other techniques in massive MIMO and reflective intelligent surface implementations. An RRF can also serve as a digital radio twin, which is a virtual representation of the radio environment that includes both geometric structure and radio propagation characteristics, enabling real-time simulation and optimization of wireless systems. It paves the way for various applications from communications to sensing in the next-generation wireless communication systems.

\end{abstract}

\section{Introduction} \label{sec:intro}

Assessment of recent statistics collected by International Telecommunication Union (ITU) indicates that global mobile broadband traffic around the world reached 1.3 zettabytes in 2024~\cite{itu_mobile_traffice}. Beyond those traditional cellular users, more and more wearable devices, machine-2-machine connections, and satellite communications are expected to be supported by the next-generation wireless communication systems with a diverse set of requirements in terms of data rates, reliability, latency, intelligence, security and energy efficiency. To support all the requirements, advanced physical layer solutions, modulation schemes, multi-access techniques, energy harvesting, edge computing, new spectral bands, integration of terrestrial and non-terrestrial communications, adoption of machine learning and artificial intelligence (AI) techniques, have been introduced and envisioned in the next-generation wireless systems.

\begin{table*}[ht]
    \centering        
    \caption{Current and emerging channel modeling and channel estimation approaches.}\label{table:channel}
    \begin{tabular}{|m{0.25\linewidth} | m{0.2\linewidth} | m{0.2\linewidth} | m{0.2\linewidth} |}
      \hline
      \textbf{Approaches}   & \textbf{Applications} & \textbf{Pros} & \textbf{Cons}  \\
      \hline
      \multicolumn{4}{|c|}{\emph{Traditional approaches}}\\
      \hline
       Empirical and statistical methods~\cite{singh2012comparison,series2015propagation}  & Large scale path loss modeling and service planning & Provide mean path loss validated for mobile networks. & Limited use case and inaccuracy for real-time purposes. \\      
      \hline
      Computational EM methods~\cite{bondeson2012computational,sullivan2013electromagnetic} & Antenna design or near-field simulations. & Effective and accurate for the use cases.
       & High computational overhead. Ineffective in general scenarios. \\
      \hline
      Pilot-based channel estimation~\cite{omar2008performance} & Real-time channel estimation for channel pre-coding and symbol recovery. & Real-time processing in OFDM and OFDMA systems. & Introduces communication and computational overhead, as well as latency.\\
      \hline
      RT methods~\cite{7152831,10556262} & Radio map construction and spectrum planning. & Accurate and flexible in various environments. & RT has high computational overhead. \\       
       \hline
      \multicolumn{4}{|c|}{\emph{Radiance field approaches}}\\
      \hline
       
       Computer vision and radiance field approaches~\cite{10684152,chen2024rfcanvas,zhao2023nerf2,zhang2024rf,11044513} & Radio map construction, spectrum planning, and real-time channel rendering. & Accurate channel representation with spatial information.  & Complex implementation pipeline and high computational overhead. \\
       \hline
    \end{tabular}
    \label{tab:channel_models}
\end{table*}

Massive multiple-input multiple-output (MIMO) is one of the key advanced technologies to enable a wider transmission bandwidth, a higher data rate, and more mobile broadband connections in 5G and beyond wireless communications systems. Built on MIMO, new research areas such as holographic antenna, fluid antenna, and reflective intelligence surface (RIS), have been widely studied recently. One of the major challenges in fully exploiting MIMO is obtaining accurate channel state information (CSI). Although channel modeling and channel estimation have been widely studied, the traditional empirical and statistical approaches provide radio propagation models, such as Okumura-Hata~\cite{singh2012comparison} and COST 231 models~\cite{series2015propagation}, typically for large-scale planning. The computational EM methods offer high accuracy for applications such as antenna design or near-field simulations~\cite{bondeson2012computational,sullivan2013electromagnetic}. These methods have limited use cases while only providing accumulated and limited CSI. 
Pilot-based channel estimation is widely applied for real-time applications in Orthogonal Frequency-Division Multiplexing (OFDM) and Orthogonal Frequency-Division Multiple Access (OFDMA) systems \cite{omar2008performance}. However, when scaled to massive MIMO systems, it introduces significant communication and computing overhead that increases rapidly with the array size, which also exacerbates the difficulty of implementing full-digital beamformers. Such costs may not be acceptable in the next-generation wireless communication systems. 
{A promising alternative is to leverage explicit spatial channel information to guide hybrid or analog massive MIMO systems. This approach involves acquiring comprehensive Channel State Information (CSI), which we refer to as \textit{Spatial-CSI}. This model can be considered as an extended double-directional radio channel composed of multipath components (MPCs), specifically including their pathloss, Angle of Arrival (AoA), Angle of Departure (AoD), and Time of Flight (ToF), with the option to include additional spatial characterizations. 
For example, the full propagation path geometry from the transmitter (Tx) to an arbitrary receiver (Rx) position. Based on this full understanding of the radio propagation environment, it is possible to utilize both line-of-sight (LoS) and non-LoS (NLoS).
Such extended characterizations are crucial for 6G applications. For instance, with RIS, full propagation geometry is required to determine optimal placement and its interactions with the Tx. Similarly, for electromagnetic (EM) environment reconstruction and calibration, it is necessary to identify which objects/materials interacting with which MPCs and subsequently perform backward optimization of the material parameters.
Ray tracing (RT) has been introduced for radio propagation modeling, where the spatial information can be preserved~\cite{7152831,10556262}. However, RT for massive MIMO requires extremely high computing resources and modeling cost.

This article introduces radio radiance field (RRF), which characterizes the spatial distribution, intensity, and directionality of radio signals as they propagate through an environment. Radiance fields are typically used in computer vision to model how light interacts with objects within the 3D scene, capturing color and intensity from various viewpoints to enhance high-quality rendering.
In contrast to traditional wireless RT, RRFs require only raw visual and radio data (Spatial-CSI), rather than a fine-grained mesh model that demands delicate material-based segmentation and high geometric accuracy of the scene. Moreover, RT must simulate arbitrary Tx and Rx configurations, including position, orientation, and antenna gain, which makes real-time execution or exhaustive result storage computationally prohibitive. By comparison, each RRF is trained for a fixed Tx configuration, consistent with many practical implementations, while still supporting arbitrary Rx configurations. The compact representation and fast inference of RRFs make them significantly more practical than RT. Nonetheless, RRFs can be interpreted as a post-processing stage of RT, effectively compressing the full set of RT results into a concise, data-driven model.

Although radiance fields have been developed in CV, it is still challenging to adapt this idea to radio frequency domain. The primary difficulty lies in the difference between how light and radio waves interact with objects. RRF models require very different radio-object interaction models because radio waves possess wavelengths that are comparable in scale to common objects. This difference necessitates two major modeling adjustments: the propagation tends to be dominated by specular reflection, demanding a highly directional model; and diffraction occurs much more frequently, requiring proper incorporation into the RRF framework. Second, the existing approaches with RT in CV applications have high computational complexity that cannot be afforded by a communication system. 
To address these challenges, a pioneering scheme is developed to provide a seamless cooperation between the representation structure, the rendering pipeline, and the training strategies by utilizing explicit environmental representation from CV, in conjunction with radio domain information to represent the RRF.

Upon addressing these challenges, RRF can provide a viable solution for accurate real-time Spatial-CSI estimation with ultra-low communication and computational overhead on transmitting or receiving devices. It also enables the design and optimal implementation of various physical (PHY)-layer focused wireless communication technologies, such as massive MIMO beamforming, RIS design, PHY-layer security, as well as integrated sensing and communication (ISAC).

\section{Current and Emerging Channel Modeling and Channel Estimation Approaches}\label{sec:state-of-the-art}

Radio propagation modeling has been extensively studied over the decades. Empirical and semi-empirical models, such as the Okumura-Hata and COST 231~\cite{singh2012comparison}, are widely used for predicting approximate path loss in large-scale environments. These models are often based on practical measurement data in the service area. Some of the approaches also rely partially on deterministic models including RT and Ikegami model. Using only a few parameters, these channel models are simple. They have also been extensively validated for mobile networks in macro cellular environment. Similarly, empirical and semi-empirical path loss models are available for indoor environment~\cite{series2015propagation}. However, the empirical and semi-empirical channel models are intended for estimating only coarse propagation parameters based on inputs such as antenna heights, distance between Tx and Rx, and statistical shadowing effects. Therefore, these models are inadequate for massive MIMO applications in the next-generation wireless systems. Computational EM simulations utilize acceleration methods such as the finite-difference time-domain method~\cite{
bondeson2012computational,sullivan2013electromagnetic}.
Channel modeling has advanced well beyond early empirical approaches. Modern techniques now include geometry-based stochastic models, standardized frameworks such as 3GPP TR 38.901, and hybrid methods that integrate RT with machine learning. These models capture spatial, temporal, Doppler, and polarization characteristics that are critical for massive MIMO systems. However, they are often tailored either to specific applications such as antenna design and near-field simulations, or to large-scale system design and network planning.

Real-time channel estimation is usually achieved by using pilot-based channel estimation, especially in OFDM and OFDMA systems. For example, Long Term Evolution (LTE) OFDMA down-link systems use pilot symbols, which are cell-specific reference signals, inserted in both time and frequency. These pilot symbols provide an estimate of the channel at given locations within a subframe~\cite{omar2008performance}. Through interpolation, such as linear, piece-wise cubic and spline, it is possible to estimate the channel at non-pilot locations for different antenna configurations. Pilot-based channel estimation has demonstrated its efficacy owing to the pervasive implementation of OFDM and OFDMA systems. However, the pilot sequences add communication overhead as well as computational overhead to wireless systems. Even worse, this feedback process adds more communication overhead in frequency division duplex (FDD) systems that operate on different bands. 

More recently, RT offers a potential solution by approximating radio waves using ray concepts. It models radio waves as rays to simulate complex RF interactions such as reflection, diffraction, and scattering~\cite{7152831}. While RT is accurate and flexible in various use cases, it requires high computational complexity as well as precise representation of the environment, e.g., 3D modeling and surface properties. Therefore, RT is not optimal for more general and real-time channel modeling in wireless communication systems, especially for massive MIMO use cases. Meanwhile, neural network methods provide a more efficient approach for channel modeling. These approaches use broader environmental information as input to generate outputs such as CSI or {path loss} maps. However, while these approaches are fast with acceptable accuracy, their use cases are usually limited within the domain defined by the training dataset.

\section{Representing Wireless Channel with RRF}~\label{sec:channel_current}

\subsection{The Radiance Field and RRF}
Radiance field as a 3D scene reconstruction method has been widely studied in CV. These representations rely on capturing the optical radiance emitted from object surfaces towards the camera. An optical radiance is usually modeled as a group of rays passing through each pixel to the camera. Since wireless channels can also be modeled by a group of MPCs arriving at the Rx from different AoAs, recent research has introduced radiance into the RF domain. For example, RayProNet~\cite{10684152} and RFCanvas~\cite{chen2024rfcanvas} integrate environmental information as input and uses an explicit ``light probe'' as the radio radiance field voxel. However, they are limited to producing only 2D path loss maps, without offering more detailed information.
Meanwhile, NeRF$^2$~\cite{zhao2023nerf2} reconstructs a ``squared'' RRF that considers both ``camera'' and ``light source'' locations as inputs. However, its spectra, generated using comparing the conventional beamforming (CBF) at 915 MHz with linear scale transformation and per-view normalization, suffer from strong interference, inconsistencies across views, lack of geometric information, and distance-dependent decay. Moreover, NeRF$^2$ has high computational complexity which requires hours of training. Its CSI rendering time is also relatively long, thus cannot be applied in real-time massive MIMO applications. More recently, RF-3DGS~\cite{zhang2024rf} and  WRF-GS~\cite{11044513} leverage 3D Gaussian Splatting (3DGS) to represent RRF with higher accuracy and faster rendering.

\begin{figure}[ht!]
    \centering
    \boxed{\includegraphics[width=0.9\linewidth]{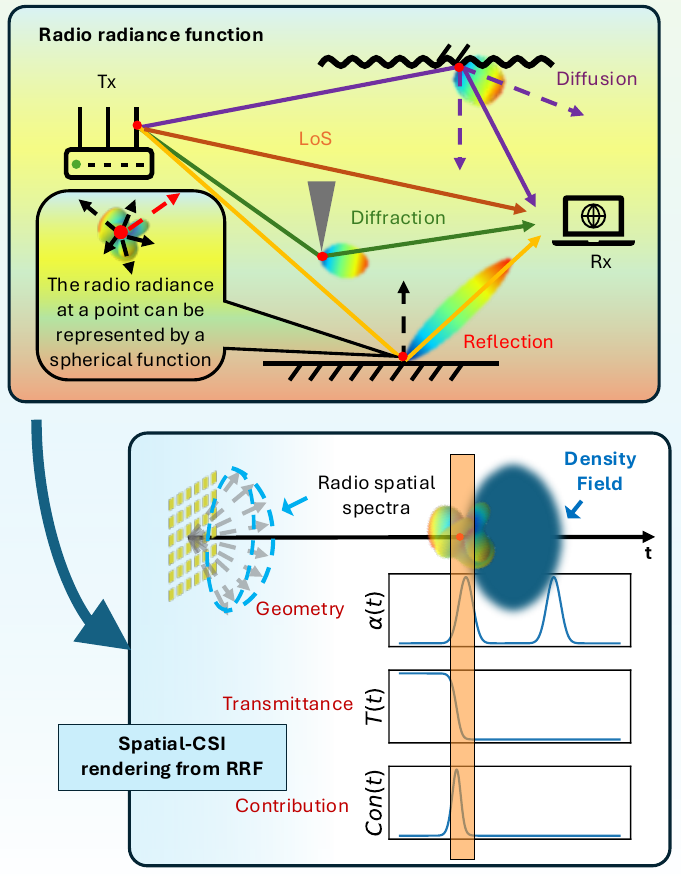}}
    \caption{Representing \emph{Spatial-CSI} with RRF.}
    \label{fig:RRF_concept}
\end{figure}

\begin{figure*}[ht]
    \centering
    \boxed{\includegraphics[width=0.95\linewidth]{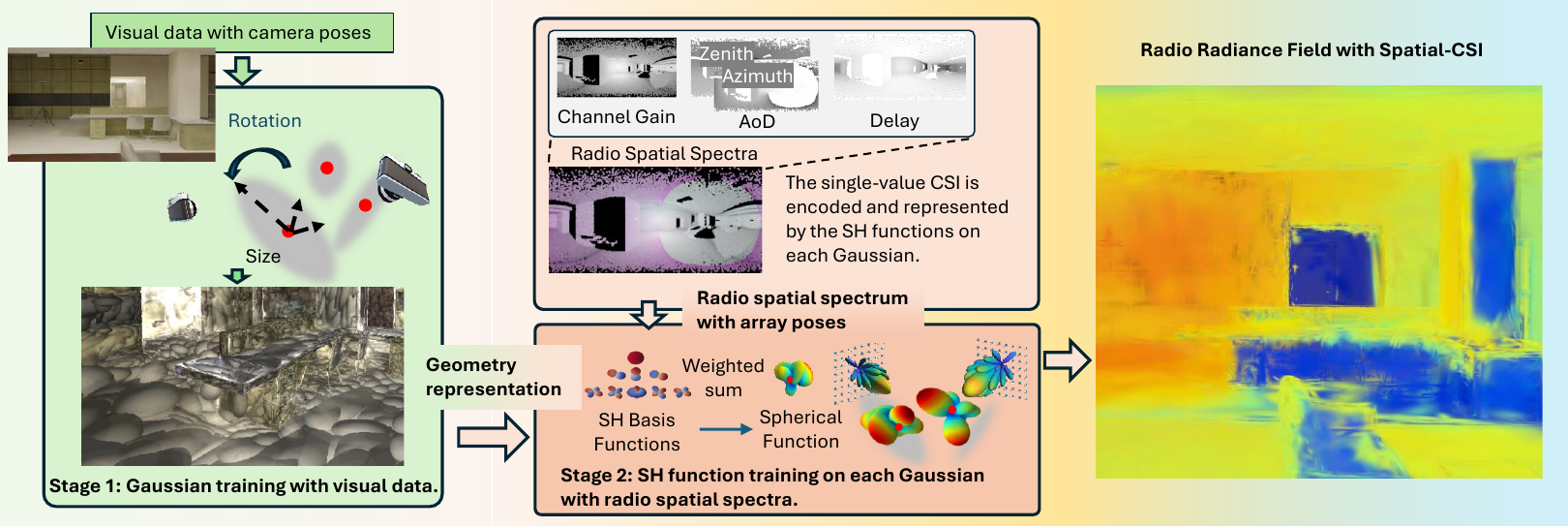}}

    \caption{RRF reconstruction using RF-3DGS pipeline.}
    \label{fig:RRF}
    \vspace{-4mm}
\end{figure*}

As shown in Fig.~\ref{fig:RRF_concept}, an RRF representation depends on an RRF function that takes into arguments of the radio radiance (3D coordinate and the radiating direction), and the geometry of the environment. Technically speaking, the radio radiance can be represented by a spherical function that describes the MPCs originating or retransmitting from this point at arbitrary direction, given a point in an environment. A specific point and direction can define the geometry of a ray in 3D space. If an Rx lies on this ray, and the ray is within the Rx's field of view (FoV) with no obstacles, the Rx can receive this MPC. Therefore, an RRF can be defined as a model that describes how radio waves are emitted from every point in a 3D space toward the entire spherical domain (the full $4\pi$ steradians), given a bounded environment and the Tx configuration. 
The key differences between radiance fields in CV and RRF lie in the underlying physics. While CV radiance fields assume optical wavelengths and physically-based rendering, RRF must model EM propagation, including diffraction, reflection, and material absorption. 
Note that unlike traditional wireless RT which separately computes specular and diffuse reflections, RRF models the total reflected energy, which is the radiance emitted from the object, as a unified representation. In doing so, it inherently combines both reflection components into how the object appears under radio frequency.
Furthermore, RRF integrates radio-specific parameters such as polarization and frequency-dependent attenuation, which are absent in CV radiance fields. Specifically, RRF handles blockages with a density field function, where zero indicates free space, low values indicate translucent objects, and high values indicate solid object surfaces. Geometry of the environment is required to get the density field function. Nonetheless, reconstructing the interior of solid objects is unnecessary for RRF representations.

\subsection{Spatial-CSI Rendering in RRF}
The wireless channel between a Tx and a Rx can be rendered as a radio spatial spectrum, given a well-trained RRF representation and the current Rx configuration.
Unlike traditional MIMO CSI measurements, which typically capture the channel frequency response at baseband between each Tx and Rx array element in vector or matrix form, the radio spatial spectrum represents the CSI of a set of MPCs across different AoAs, as illustrated in Fig.~\ref{fig:RRF}. Specifically, Spatial-CSI records information equivalent to a 3D channel tensor, describing the frequency response between each Tx and Rx array element. The radio spatial spectrum unfolds this tensor into the AoA domain, enabling efficient querying of Spatial-CSI for dominant MPCs and supporting higher-level applications. 
Spatial-CSI can be queried at uniformly distributed AoAs within the array’s FoV. While most queried AoAs may not correspond to strong MPCs, the illustration assumes the presence of MPCs at each AoA. Collectively, these queried Spatial-CSI values form the radio spatial spectrum, which provides an accurate depiction of the wireless channel between Tx and Rx by incorporating both radio and spatial information.
Techniques such as the alpha-blending rendering technique can be used for querying at each AoA. For example, a rendering ray can be launched along the AoA being queried. In this way, the 3D position can be simplified to a 1D coordinate on the ray for convenience.
Note that only the radio radiance of the first encountered surface point significantly influences the received Spatial-CSI. For example, as shown in the highlighted strip in the lower half of Fig.~\ref{fig:RRF}, when the rendering ray intersects a high-density field, the transmittance drops sharply from 1 to 0. As a result, the portion of the ray behind the intersection has little contribution to the RRF. Similarly, the portion of the ray in free space before the intersection also contributes little due to their zero density. Note that while current RRF models emphasize carrier frequency, they naturally extend to wideband channels. This is because the EM properties of materials and the reflection patterns determined by surface smoothness remain largely consistent, even across bandwidths as wide as 300 MHz. In contrast, the frequency selectivity typically associated with wideband channels is captured by the channel frequency response, sampled at the baseband subcarriers. This selectivity arises from the underlying physical channel impulse response, which is inherently recorded within the RRF.

\section{RRF Reconstruction}
RRF reconstruction requires a learnable representation capable of encoding the geometry and the radio radiance, while maintaining differentiability to the radio radiance function as well as the radio spatial spectrum function. In addition, the representation should be editable, concise, and easy to initialize. For example, the alpha-blending rendering equation is differentiable, allowing the loss and representation to be directly coupled. Given a well-initialized learnable RRF representation, it is then iteratively optimized through gradient descent to minimize the loss between the rendered radio spatial spectra and the training spectra. Nonetheless, the efficiency of this training process requires seamless cooperation between the representation structure, the rendering pipeline, and the training strategies. Assuming all the requirements are met, RRF reconstruction can be done as follows. Given a surface point, each spectrum contains a rendering ray that renders this point during optimization. As all these rays converge at the actual position of the surface point in 3D space, during gradient descent, only this surface point in the representation should receive an increase in density and its corresponding Spatial-CSI, as this is the optimal solution to minimize the loss functions associated with those rays. 

Fig.~\ref{fig:RRF} depicts an accurate and efficient RRF reconstruction method. Building on 3D Gaussian Splatting (3DGS), our RF-3DGS~\cite{zhang2024rf} provides a seamless cooperation between the representation structure, the rendering pipeline, and the training strategies by utilizing explicit 3D Gaussian distributions and spherical harmonic (SH) Functions to represent the RRF. As the millions of 3D Gaussians collectively define a density field function, and the SH functions define the radio radiance of discrete object points (the means of the Gaussians), the alpha-blending rendering can be directly applied to the Gaussians along each ray, without relying on complex and lossy sampling strategies. Specifically, an efficient rasterization rendering pipeline is implemented. Given the current rendering image plane and view frustum, the image plane is divided into $16\times 16$ tiles, and the view frustum is divided accordingly. For each tile, only the Gaussians within the corresponding divided view frustum are selected (following a specific criterion) and sorted based on their depth to the plane. The computation within each tile consists of parallelized threads rendering the sorted Gaussians onto each pixel. Once a pixel is saturated (reaching near-zero transmittance), meaning that further Gaussians cannot affect this pixel, the
thread rendering this pixel is terminated.

\begin{figure}[ht!]
    \centering
    \boxed{\includegraphics[width=0.99\linewidth]{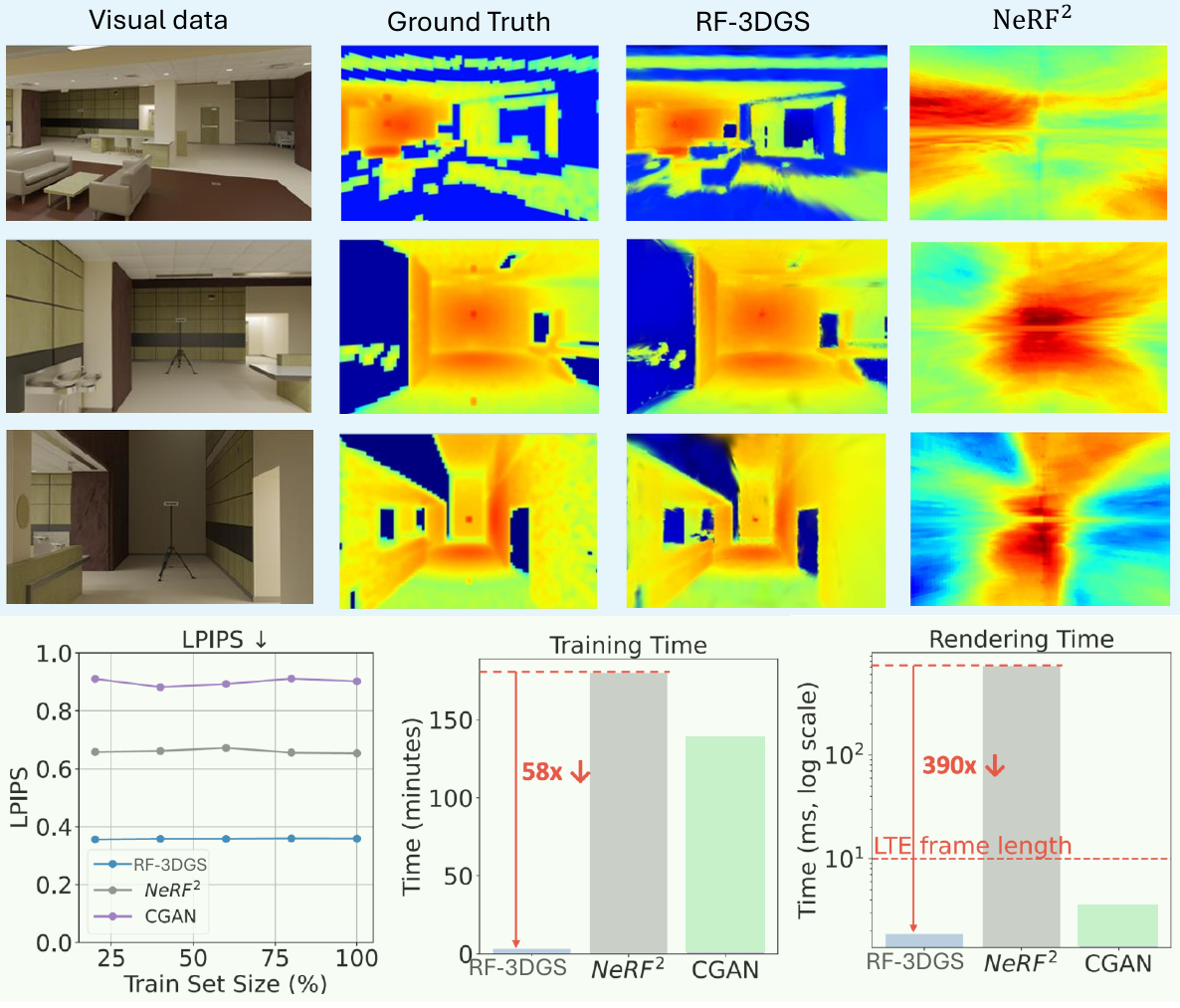}}
    \caption{Results of RRF reconstruction.}
    \label{fig:results}
    \vspace{-2mm}
    
\end{figure}

RF-3DGS applies a two-stage fusion training process by assuming access to the radio spatial spectra and the visual data simultaneously. The first stage trains the Gaussians in multiple phases using images with different resolutions, respectively. This stage of training enables the global geometry representation. In the second stage, the CSI-encoded SH functions and the basic density of each Gaussian are trained using the collected radio spatial spectra, while freezing the geometry representation from the first stage. The new CSI-encoded SH functions can be interpreted as the objects being re-illuminated by the given Tx radio, with multiple channels corresponding to different Spatial-CSI properties. {As demonstrated in Fig.~\ref{fig:results}, in terms of  learned perceptual image patch similarity (LPIPS, the lower the better), which is used to judge the perceptual similarity between two images, RF-3DGS can reconstruct RRF more accurately compared with other related methods such as state-of-the-art NeRF$^2$ and the conditional generative adversarial network (CGAN)~\cite{zhao2023nerf2}. More importantly, both the training and rendering efficiency are much improved (58$\times$ and 390$\times$, respectively). Especially the rendering efficiency has reached real-time capability, at 2 ms or lower. }

\section{RRF Applications}\label{sec:applications}

\subsection{Digital Radio Twin}

Digital radio twin would be one of the major applications delivered by RRF, owing to its accurate and real-time inference and explicit geometric representation of Spatial-CSI. As demonstrated in~\cite{zhang2024rf}, the radio spectra in a well-implemented RRF align well with the equirectangular photographs taken from the same positions, in addition to its strong Spatial-CSI metrics. Demonstration of digital radio twin implementations is shown in Fig.~\ref{fig:digital_radio_twin}. In comparison, field measurements are usually performed for the creation and validation of a digital radio twin. For example, MPCs such as path loss, delay, AoD and AoA can be captured between each Tx and Rx array element pair using a synchronized Tx-Rx system that generates precise channel impulse response~\cite{papazian2016calibration}. Note that this particular field measurement process could suffer from two limitations: a limited Rx vertical FoV and an estimation on dominant MPCs only. As a result, a digital radio twin created from this approach can only produce little spectrum information beyond those MPCs. The proposed RRF, especially when using the RF-3DGS RRF reconstruction method, can generate a highly-reliable digital radio twin within a few minutes.

\begin{figure}[ht!]
    \centering
    \boxed{\includegraphics[width=0.95\linewidth]{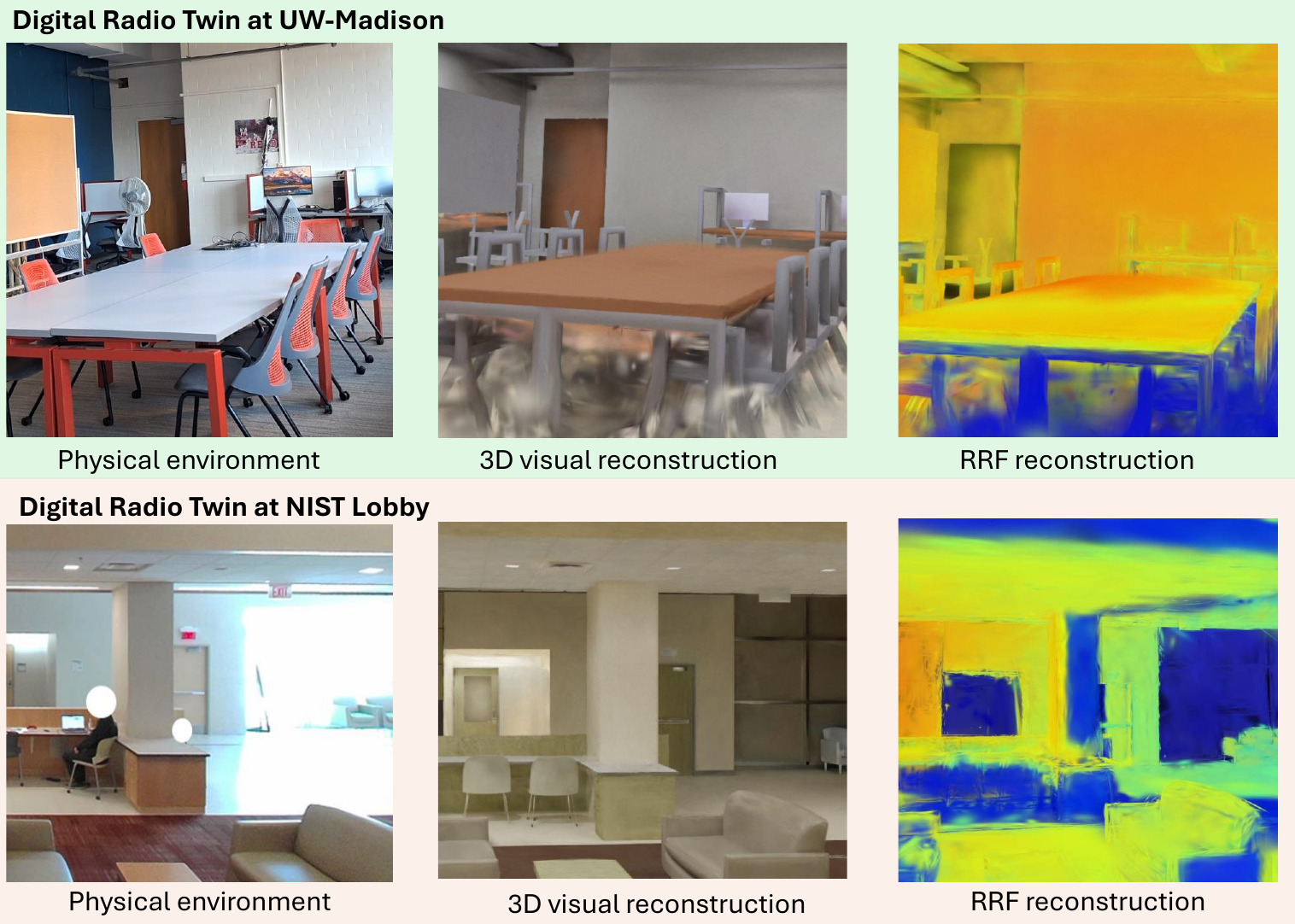}}
    \caption{Application of RRF in digital radio twins.}
    \label{fig:digital_radio_twin}
    \vspace{-8mm}
\end{figure}

\begin{figure*}[ht]
    \centering
    \boxed{\includegraphics[width=0.95\linewidth]{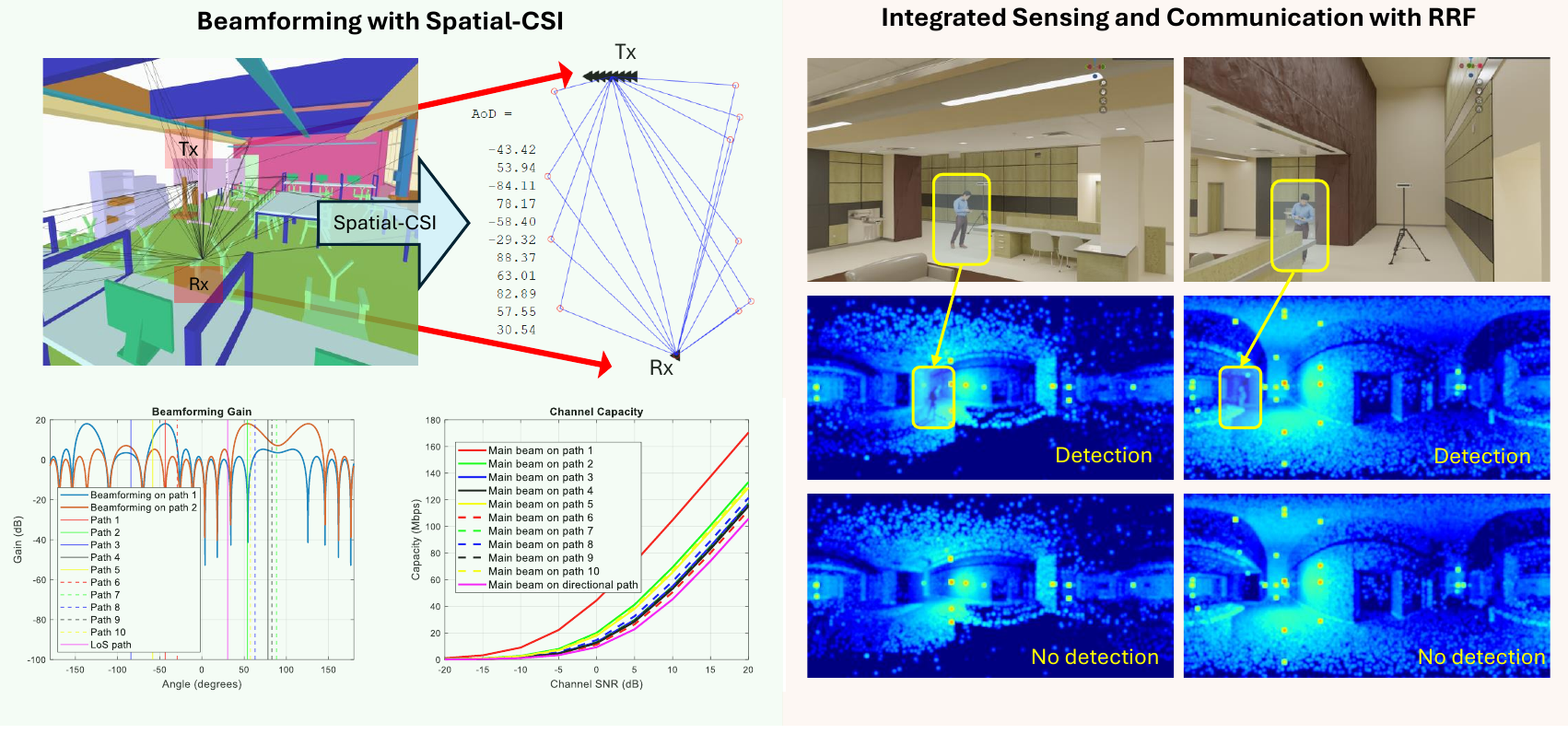}}
    \caption{Beamforming with Spatial-CSI and ISAC using RRF.}
    \label{fig:RRF_app}
    \vspace{-4mm}
\end{figure*}

\subsection{Facilitate Communication and Sensing with RRF }
Owing to the fundamental ability to model propagation between arbitrary points, RRF can greatly enhance the optimization and implementation of beamforming techniques in massive MIMO as well as RIS. To achieve the two-way channel required by RIS, two queries are generally needed. One is to render the incident channel spectrum by treating the Tx location as the source and generating a radio spectrum (or directional field) at the RIS's designated pose. The other is to render the reflected channel spectrum at the target Rx pose.
For example, as shown in the left half of Fig.~\ref{fig:RRF_app}, the steering vector can be optimized straightforwardly along each path given the AoD, gain, delay and polarization, to reach the desired channel capacity while meeting the latency requirement. More advanced beamforming techniques, such as delay alignment modulation~\cite{lu2022delay}, can also benefit directly from Spatial-CSI. Moreover, RRF can play an important role in ISAC applications. For example, as demonstrated in the right half of Fig.~\ref{fig:RRF_app}, when using CSI for object detection, the RRF plot clearly shows the target when the testing object (a person in this case) enters the scene. Comparing to a traditional visual detection method, e.g., by using a camera, RRF generally has its advantage in a much larger FoV that covers the entire environment. Note that this demo does not yet generalize the sensing module for purposes such as localization or human posture perception. To add such functionalities, one may need to harness information from multiple devices at different positions for estimation; and set certain parameters of the human model as learnable, importing this to the digital radio twin, and using gradient descent to learn those parameters.

\section{Future Research Directions}\label{sec:future}

\subsection{Active Learning and Bayesian Optimization for Efficient Sampling}
To accurately reconstruct an RRF, it is essential to take wireless channel samples in the field. However, the process of obtaining sufficient wireless samples presents several challenges. Measurement campaigns require extensive time, specialized equipment, and careful calibration to ensure consistency and accuracy. Additionally, environmental factors such as interference, obstacles, and dynamic conditions further complicate data collection. As a result, an efficient sampling strategy is necessary to minimize measurement overhead while maximizing reconstruction accuracy. In CV, recent research has explored methods to strategically select fewer but more informative samples for visual reconstruction. These approaches leverage active learning and Bayesian optimization to guide sampling based on the uncertainty of the learned model, prioritizing regions that contribute the most to reconstruction quality. Similar principles can be applied to wireless measurements, where an optimal sampling strategy should aim to balance spatial coverage and information gain. Instead of uniformly distributing samples, an adaptive strategy can be employed to focus on areas with higher signal variations or uncertainty in the learned RRF model.

One major research direction is the design of uncertainty-driven sampling policies. By using Gaussian processes or neural network-based uncertainty estimation, measurement locations can be dynamically selected to refine the RRF with minimal samples. Another challenge is dealing with non-stationary environments where the channel characteristics evolve over time. Incorporating temporal factors into the sampling strategy, such as reinforcement learning-based adaptive sampling, could help efficiently track variations in the wireless environment.
Furthermore, the integration of priors from RT simulations, 3D maps, or historical data could reduce the number of required measurements while improving accuracy. Leveraging hybrid approaches that combine model-based and data-driven techniques may further enhance sampling efficiency, making it feasible to reconstruct RRF with fewer yet highly informative measurements.

\subsection{Visual Representation and End-to-End Learning}
Our current work leverages the advances from CV to create an efficient visual representation in the first stage, then use sparse radio samples for the Spatial-CSI reconstruction \cite{zhang2024rf}. One limitation is that due to different objectives of visual and radio tasks, such visual representation may not be ideal for wireless channel modeling. For example, although 3DGS provides fast training and rendering speed, the geometry is loosely represented by Gaussian ellipsoids, which cannot reflect the true object surface that is critical for wireless channel propagation. Besides, visual signals are not sensitive to distance changes. For example, viewing an object from 1 or 2 meters apart does not incur msignificant changes in brightness (in terms of pixel strength). However, wireless signals should vary significantly for different distances. This calls for a more consistent geometry representation, as well as a more advanced training pipeline to incorporate wireless-specific information into the future RRF.

\subsection{RRF in the Wild}
Producing RRF in unconstrained environments, such as urban areas or open landscapes, introduces new challenges compared to controlled indoor settings. Outdoor wireless propagation is affected by dynamic factors like weather, moving obstacles, and environmental clutter, making it difficult to ensure consistent measurements. Unlike structured environments, where sampling can be planned systematically, real-world conditions demand adaptive and robust reconstruction techniques.
One of the key challenges is how to deal with occlusions and NLoS regions, where signals undergo diffraction, reflection, and scattering. Incorporating priors from 3D maps, RT, or spatial statistics can help fill in missing data and improve reconstruction accuracy. Additionally, temporal variations in outdoor settings require strategies that can update the RRF over time, such as reinforcement learning-based sampling or online adaptation.
Another major challenge is adapting to environmental changes, such as shifting traffic patterns, seasonal foliage variations, or evolving urban infrastructure. A static RRF model may become outdated quickly, requiring periodic re-sampling and real-time updates. Methods that integrate continual learning, sensor fusion, and predictive modeling can help track and compensate for these changes, ensuring a more reliable and adaptive RRF reconstruction in the wild.

\section{Conclusion}\label{sec:conclusion}

This article introduced RRF as a new frontier of spatial wireless channel representation. Wireless channel represented by RRF consists of both geometric and radio information. Rooted from radiance field reconstruction methods, RRF mainly captures the geometric information from similar CV techniques. However, it is not straightforward to reconstruct RRF just using CV approaches because of the vastly different properties presented in EM waves compared with visible light as well as its distinct path-loss characteristics. This article hence introduced a practical RRF reconstruction pipeline that addresses these issues. Owing to the high accuracy and real-time rendering efficiency, an RRF with Spatial-CSI can be implemented directly to enable various massive MIMO techniques such as beamforming, delay-alignment modulation, and RIS implementations. An RRF also serves as a digital radio twin, which lays a solid foundation for various applications from communications to sensing in the next-generation wireless systems.

\renewcommand\refname{Reference}
\bibliographystyle{IEEEtran}
\bibliography{References}

\section*{Biography}\label{sec:bio}

\begin{IEEEbiography}{Haijian Sun} [Senior Member, IEEE] (hsun@uga.edu) is an Assistant Professor in the School of Electrical and Computer Engineering at The University of Georgia. He obtained Ph.D. degree in the Department of Electrical and Computer Engineering from Utah State University, USA, in 2019. His current research interests include vehicular communication, wireless communication for 5G and beyond, machine learning at the edge, cyber security, IoT communications, wireless systems, and optimization analysis. Dr. Sun directs the ESI Wireless Lab at The University of Georgia and has published extensively in the field of wireless communication. He is a recipient of the Best Paper Award from IEEE ISICN 2024. Dr. Sun is currently serving on the editorial boards for IEEE Communications Magazine and IEEE Network. 
\end{IEEEbiography}

\begin{IEEEbiography}{Feng Ye} [Senior Member, IEEE] (feng.ye@wisc.edu) is currently an assistant professor in the Department of Electrical and Computer Engineering, University of Wisconsin-Madison, Madison, WI, USA. He obtained a Ph.D. degree in Electrical \& Computer Engineering from the University of Nebraska-Lincoln (UNL), NE, USA, in 2015. His research interests include wireless communications and networks, artificial intelligence in networks, cyber security, and big data analytics. He is currently a column editor of IEEE Wireless Communications; an associate editor of IEEE Transactions on Vehicular Technology and IEEE Internet of Things Journal.
\end{IEEEbiography}

\end{document}